\documentclass[prl,twocolumn,floatfix,superscriptaddress]{revtex4-1}
\usepackage{amsmath}
\usepackage[dvips]{graphicx}
\usepackage{graphicx}
\usepackage{bm}
\usepackage{hyperref}

\newcommand{\ecm}{\ensuremath{e {\cdotp} {\rm cm}}}
\newcommand{\eEDM}{{\em e}EDM}
 \newcommand{\Eeff}{\mathcal{E}_\mathrm{eff}}
\newcommand{\Bvec}{\vec{\mathcal{B}}} 
\newcommand{\Evec}{\vec{\mathcal{E}}} 

\newcommand{\de}{d_\mathrm{e}}

\begin{document}
   \title{Interference between $E1$ and $M1$ transition amplitudes on the $H$ to $C$ transition in ThO}
\author{A.N.\ Petrov}\email{alexsandernp@gmail.com}
\author{L.V.\ Skripnikov}
\affiliation
{National Research Center ``Kurchatov Institute'' B.P. Konstantinov Petersburg
Nuclear Physics Institute, Gatchina, Leningrad district 188300, Russia}
\affiliation{Department of Physics, St. Petersburg University,7/9 Universitetskaya nab., St. Petersburg, 199034 Russia}


\begin{abstract}
Calculations of Stark interference between $E1$ and $M1$ transition amplitudes on the $H^3\Delta_1$ to $C^1\Pi$ transition in ThO is performed. Calculations are required for estimations of systematic errors in the experiment for electron electric dipole (\eEDM\ ) moment search
due to imperfections in laser beams used to prepare the molecule and read out the \eEDM\ signal.%
\end{abstract}
\maketitle
%
The current limit for the electron electric dipole moment (\eEDM) (ACME II experiment), 
$|\de|<1.1\times 10^{-29}$ \ecm\ (90\% confidence),
was set by measuring the spin precession  
of thorium monoxide (ThO) molecules in the metastable electronic $H^3\Delta_1$ state 
\cite{ACME:18}. Previous experiment (ACME I) used ThO to place a limit $|\de|<9\times 10^{-29}$ \ecm\ (90\% confidence).
In both experiments the measurements were performed on the ground rotational level which has two closely-spaced $\Omega$-doublet levels of opposite parity.
It was shown that due to existence of closely-spaced $\Omega$-doublet levels the experiment on ThO is very robust against a number of systematic effects \cite{DeMille2001, Petrov:14, Vutha2009, Petrov:15,Petrov:18}.
In ACME I the state preparation and the spin precession angle, $\phi$, measurement is performed by optically pumping the  $H^3\Delta_1 \rightarrow C^1\Pi$
transition with linearly polarized laser beam \cite{Baron2013}, whereas In ACME II
the  $H^3\Delta_1 \rightarrow I^1\Pi$ transition was used.
The transitions to ground rotational level of $C^1\Pi$ (ACME I) or $I^1\Pi$ (ACME II) which have similar to $H^3\Delta_1$ $\Omega$-doublet structure (see below) is used. 
Searching for systematic errors is an important part of the \eEDM~ search experiment. It was found that one of the systematic errors in the ACME I experiment \cite{Baron2013} are due to 
Stark interference between $E1$ and $M1$ transition amplitudes on the $H^3\Delta_1$ to $C^1\Pi$ transition in ThO in laser beams used to prepare the molecule and read out the \eEDM~ signal \cite{ACME:17}. 
The similar systematic error related to  $H^3\Delta_1$ to $I^1\Pi$ transition is expected for the ACME II experiment.


The measurement of spin precession is repeated under different conditions which can be characterized by binary parameters being switched from $+1$ to $-1$. 
The three primary binary parameters are $\tilde {\cal N}$, $\tilde {\cal E}$, $\tilde {\cal B}$.
${\cal \tilde{N}} {=} +1(-1)$ means that the measurement was performed for lower (upper) $\Omega$-doublet level of $H^3\Delta_1$.
$\tilde {\cal E} = {\rm sgn}(\hat{z}\cdot\Evec)$ and $\tilde {\cal B} = {\rm sgn}(\hat{z}\cdot\Bvec)$ define the orientation of the external static electric and magnetic
fields respectively along the laboratory axis $z$.
The measured precession angle $\phi$ can be represented as \cite{ACME:17}

\begin{eqnarray}
\nonumber
\phi({\cal \tilde{N}},{\cal \tilde{E}},{\cal \tilde{B}}) = \phi^{nr} 
+{{\cal \tilde{B}}}\phi^{{\cal {B}}}
+{{\cal \tilde{E}}}\phi^{{\cal {E}}}
+{{\cal \tilde{E}}{\cal \tilde{B}}}\phi^{{\cal {E}}{\cal {B}}} \\
\label{phiNEB}
+{{\cal \tilde{N}}}\phi^{{\cal {N}}}
+{{\cal \tilde{N}}{\cal \tilde{B}}}\phi^{{\cal {N}}{\cal {B}}}
+{{\cal \tilde{N}}{\cal \tilde{E}}}\phi^{{\cal {N}}{\cal {E}}}
+{{\cal \tilde{N}}{\cal \tilde{E}}{\cal \tilde{B}}}\phi^{{\cal {N}}{\cal {E}}{\cal {B}}},
\end{eqnarray}
where notation $\phi^{S_1,S_2...}$ denotes a component which is odd under the switches $S_1,S_2,...$; $\phi^{nr}$
is a component which is even (unchanged) under any of the switches. The \eEDM\ signal is extracted from the ${{\cal \tilde{N}}{\cal \tilde{E}}}$-correlated 
component of the measured phase, $\phi^{{\cal {N}}{\cal {E}}}=\de\Eeff \tau$ \cite{Baron2013}, where
$\Eeff=79.9~ {\rm GV/cm}$ \cite{Skripnikov:13c,Skripnikov:15a,Skripnikov:16b} 
(see also Ref.\cite{Fleig:16})
is the effective electric field acting on 
the
\eEDM\ in the molecule, $\tau$ is interaction time.
In case of an ideal experiment only the interaction with 
the
\eEDM\, $\de\Eeff$, contributes
to $\phi^{{\cal {N}}{\cal {E}}}$.
However, as is stated above, Stark interference between $E1$ and $M1$ transition amplitudes
leads to an additional contribution to $\phi^{{\cal {N}}{\cal {E}}}$ and gives rise to systematic errors in the \eEDM\ measurement.
The aim of the present work is to consider this effect for ACME I experiment.
The theory can be applied to ACME II experiment as well but require more complicated
electronic structure calculation (see section \ref{ElSt}) and will be performed in forthcoming
work.
\section{Electronic structure calculations}
\label{ElSt}

For molecular calculations the following matrix elements are required:
\begin{eqnarray}
 \label{dp}
   D_+^{XC} &=& \langle C^1\Pi_1   | \hat{D}_+ | X^1\Sigma_0 \rangle, \\
 \label{gp}
   G_{+}^{XC} &=& \langle C^1\Pi_1  | \hat{L^e}_+ + g_e\hat{S^e}_+ |X^1\Sigma_0 \rangle, \\
 \label{jp}
   J_{+}^{XC} &=& \langle  C^1\Pi_1 | \hat{J^e}_+ | X^1\Sigma_0 \rangle, \\   
 \label{dz}
   D_{||}^{HC} &=& \langle  C^1\Pi_1  | \hat{D}_z | H^3\Delta_1 \rangle, \\
 \label{gp}
   G_{||}^{HC} &=& \langle  C^1\Pi_1  | \hat{L}_z + g_e\hat{S}_z | H^3\Delta_1  \rangle,
\end{eqnarray}
where $g_e=2.0023193$ is the free-electron \emph{g}-factor, $\hat{D}$ is the dipole moment operator,
$\hat{J^e}$, $\hat{L^e}$, $\hat{S^e}$ are the total, orbital and spin electronic angular moment operators; $\hat{D_+} = \hat{D}_x + i\hat{D}_y$ and the same is for other vectors.

To calculate these matrix elements two basis sets were used.
The first basis set, LBas, includes 27s, 25p, 23d, 6f, 3g, 2h and 1i (contracted for f, g, h and i-harmonics) Gaussian functions on Th and can be written in the form: (27s,25p,23d,15f,10g,10h,5i)/[27s,25p,23d,6f,3g,2h,1i]. LBas corresponds to the  aug-cc-pVTZ basis set, (11s,6p,3d,2f/[5s,4p,3d,2f], \cite{Dunning:89, Kendall:92} for oxygen. The MBas basis set was also used: (25s,22p,21d,14f,10g)/[25s,22p,21d,5f,3g] for Th and aug-cc-pVTZ for O \cite{Dunning:89, Kendall:92}. The technique of constructing natural basis sets, developed in~\cite{Skripnikov:13a} was used for constructing contracted $ f-i $ functions. $1s..4f$ electrons of thorium were excluded from the explicit electronic calculations using the generalized relativistic effective core potential in its semilocal formulation~\cite{Mosyagin:10a,Mosyagin:16}.

The transition matrix elements were calculated using the linear response coupled cluster with single and double cluster amplitudes method, LR-CCSD~\cite {Kallay:5}.
20 electrons  ($6s6p6d7s$ of Th and $1s2s2p$ of O) were included in the main correlation calculation which was performed using the LBas basis set. To calculate the correction on the correlation of the $5s5p5d$ outer-core (OC) electrons of Th, the MBas basis set was used. All calculations were performed with R(Th--O) = 3.5107 a.u. which corresponds to the equilibrium geometry of the H$^3\Delta_1$ state.

For the calculations, the  {\sc dirac15}~\cite{DIRAC15} and {\sc mrcc}~\cite{MRCC2013} codes were used.
To calculate matrix elements (\ref{dp})-(\ref{gp}) the code developed in Refs.~\cite{Skripnikov:15b,Skripnikov:15a,Skripnikov:15d} was used.

\section{Molecular calculations}
The basis set describing the $H^3\Delta_1$ and $C^1\Pi$ states wave functions can be presented as product of electronic and rotational wavefunctions $\Psi_{H(C)\Omega}\theta^{J}_{M,\Omega}(\alpha,\beta)$. Here $\Psi_{H(C)\Omega}$ is the electronic wavefunction of the $H^3\Delta_1$ ($C^1\Pi$) state, $\theta^{J}_{M,\Omega}(\alpha,\beta)=\sqrt{(2J+1)/{4\pi}}D^{J}_{M,\Omega}(\alpha,\beta,\gamma=0)$ is the rotational wavefunction, $\alpha,\beta,\gamma$ are Euler angles, and $M$ $(\Omega = \pm 1)$ is the projection of the molecular angular momentum ${\bf J}$ on the laboratory $\hat{z}$ (internuclear $\hat{n}$) axis. For short, we will designate the basis set as $\left|H(C),J, M,\Omega\right>$.
In this paper the $\left|H,J=1,\Omega, M = \pm 1\right>$ and $\left|C,J=1,\Omega, M = 0\right>$ states which are of interest for 
the \eEDM\ search experiment are considered. 

In the absence of 
the
external electric field each rotational level splits into two sublevels,
called $\Omega$-doublet levels.
One of them is even $({\cal \tilde{P}}=1)$ and 
the
 another one is odd $({\cal \tilde{P}}=-1)$ with respect to change
 the sign of electronic and nuclear coordinates.
The states with ${\cal \tilde{P}}=(-1)^J$ denoted as $e$ and with ${\cal \tilde{P}}=(-1)^{J+1}$ denoted as $f$ are the 
linear combination of the states with opposite sign of $\Omega$:

\begin{eqnarray}
\nonumber
\left|H(C)J,{\cal \tilde{P}},M\right> = \\
\label{Pstate}
\frac{\left|H(C),J,1,M\right> \pm (-1)^J {\cal \tilde{P}} \left|H(C),J,-1,M\right>} {\sqrt{2}}.
\end{eqnarray}
The experimental values of the $\Omega$-doubling, $\Delta(J)= E(\left|e,J,M\right>) - E(\left|f,J,M\right>)$ are
$\Delta_H=+0.181\,J(J+1)$~MHz for $\left|H\right>$ and $\Delta_C = -25\,J(J+1)$~MHz for $\left|C\right>$ states correspondingly \cite{Baron2013}.

External electric field $\Evec = {\cal \tilde{E}}{\cal E}\hat{z}$ does not couple the
$\left|C,J{=}1,{\cal \tilde{P}}=-1,M{=}0\right>$ and $\left|C,J{=}1,{\cal \tilde{P}}=+1,M{=}0\right>$ states, whereas the $\left|H,J{=}1,{\cal \tilde{P}}=-1,M{=}\pm1\right>$ and $\left|H,J{=}1,{\cal \tilde{P}}=+1,M{=}\pm1\right>$
states are coupled. Neglecting the interaction between
different rotational and electronic states

\begin{eqnarray}
\nonumber
\left|H,{\cal \tilde{E}}, {\cal \tilde{N}}, M\right> = k(-{\tilde {\cal N}})\left|H,J{=}1,{\cal \tilde{P}}{=}-1,M{=}\pm1\right> \\
\label{Nstate}
 - k(+{\tilde {\cal N}}){\cal \tilde{E}}{\cal \tilde{N}}M\left|H,J{=}1,{\cal \tilde{P}}{=}+1,M{=}\pm1\right> ,
\end{eqnarray}
where

\begin{equation}
k(\pm1) = \frac{1}{\sqrt{2}}\sqrt{1 \pm \frac{\Delta_H(J{=}1)}{\sqrt{\Delta_H(J{=}1)^2 + (d_H {\cal E})^2}} },
\label{RK}
\end{equation}
$d_H = -1.612$ a.u. is the dipole moment for the
$H$ state \cite{Vutha2011, Hess2014Thesis},
${\cal E}>0$ is the magnitude of the
electric field, ${\cal \tilde{E}}$ defines direction of electric field.

The dark state (which the preparation laser does not couple to the $C$ state), $\left|H_D,{\cal \tilde{E}}, {\cal \tilde{N}}\right>$ is the initial state, for the spin precession experiment. 
Let the preparation laser polarization is exactly linear $\hat{\epsilon}_{p} = \hat{x}$.
Then, neglecting the small contribution from magnetic amplitude, the resulting initial (dark) state under ideal conditions is
\begin{equation}
\label{dark}
\left|H_D,{\cal \tilde{E}}, {\cal \tilde{N}}\right> = \frac{1}{\sqrt{2}}
\left( \left|H,{\cal \tilde{E}}, {\cal \tilde{N}}, +1\right> + \left|H,{\cal \tilde{E}}, {\cal \tilde{N}}, -1\right> \right)
\end{equation}

Then the molecules enter a spin precession region with presence of electric and magnetic fields
which produce a relative energy shift between the two Zeeman sublevels
$\left|H,{\cal \tilde{E}}, {\cal \tilde{N}}, \pm1\right>$. The final state of the molecule is
\begin{equation}
\label{finalS}
\Psi(\phi)=
\frac{1}{\sqrt{2}}
\left( e^{-i\phi}\left|H,{\cal \tilde{E}}, {\cal \tilde{N}}, +1\right> + e^{i\phi}\left|H,{\cal \tilde{E}}, {\cal \tilde{N}}, -1\right> \right).
\end{equation}

Then in a detection region $\phi$ is measured by optically pumping on the same
 $H^3\Delta_1 \rightarrow C^1\Pi$ transition with linearly polarized laser beams
with polarizations $\hat{\epsilon}_{X}$, $\hat{\epsilon}_{Y}$ determined
by azimuthal angles $\theta_X=45^\circ$, $\theta_Y=135^\circ$ (azimuthal angle
for preparation laser $\theta_p=0^\circ$). Then, neglecting the small contribution from magnetic amplitude, for $\phi \ll 1$, for exactly linear polarizations of preparation and readout lasers
one can obtain \cite{Baron2013}
\begin{equation}
\label{Asym}
{\cal A}(\Psi(\phi)) = \frac{F_X-F_Y}{F_X+F_Y} = 2\tilde{P}\phi,
\end{equation}
where $\cal{A}$ is asymmetry, $F_{X,Y}$ are the detected in the experiment fluorescence after applying the readout lasers. Then the \eEDM\ sensitive component $\phi^{{\cal {N}}{\cal {E}}}$ can be calculated as

\begin{equation}
\label{components}
\phi^{{\cal {N}}{\cal {E}}} = \frac{1}{8}\sum_{\tilde {\cal B}, \tilde {\cal N}, \tilde {\cal E}} {\cal {N}}{\cal {E}} \phi\left(   \tilde {\cal N}, \tilde {\cal B}, \tilde {\cal E} \right).
\end{equation}
Similarly, other components in Eq. (\ref{phiNEB}) can be calculated.

The laser pointing vector $\hat{k}$ and polarization can be parameterized as
\begin{equation}
\hat{k}=\cos{\varphi}\sin{\vartheta}\hat{x} + \sin{\varphi}\sin{\vartheta}\hat{y} + \cos{\vartheta}\hat{z}
\end{equation}
\begin{equation}
\hat{\epsilon} = \epsilon_x\hat{x} + \epsilon_y\hat{y} + \epsilon_z\hat{z},
\end{equation}
where
\begin{equation}
\epsilon_x = \cos{\theta}(\cos{\Theta}+\sin{\Theta}) + i\sin{\theta}(\sin{\Theta}-\cos{\Theta})
\end{equation}
\begin{equation}
\epsilon_y = \sin{\theta}(\cos{\Theta}+\sin{\Theta}) + i\cos{\theta}(\cos{\Theta}-\sin{\Theta})
\end{equation}
\begin{eqnarray}
\nonumber
\epsilon_z = \tan{\vartheta}( \cos(\theta-\varphi)(\cos{\Theta}+\sin{\Theta}) + \\
i\sin(\theta-\varphi)(\sin{\Theta}-\cos{\Theta}) ),
\end{eqnarray}
$\Theta$ is the elipticity angle.
For an ideal experiment $\Theta_{p,X,Y}=45^\circ$, $\vartheta_{p,X,Y} = 0^\circ$,
$\theta_p=0^\circ$, $\theta_X=45^\circ$, $\theta_Y=135^\circ$. Label $i=p,X,Y$ refer to preparation and readout $X,Y$ lasers.
The deviation of $\Theta, \vartheta, \theta$ from their ideal values together
with Stark interference between $E1$ and $M1$ transition amplitudes
generates a systematic error in searches for the
\eEDM\ according to \cite{ACME:17}

\begin{eqnarray}
\nonumber
\tilde{\phi}^{{\cal {N}}{\cal {E}}}=\frac{a_{M1}}{4} [ \vartheta_p^2(-2S_pc_p + \tilde{P}s_p(S_X-S_Y)) +  \\
\vartheta_X^2(S_Xc_X + \tilde{P}S_ps_X) + \vartheta_Y^2(S_Yc_Y - \tilde{P}S_ps_Y)],
\label{syseq}
\end{eqnarray}

where, $a_{M1}$ is ratio of $M1$ and $E1$ amplitudes, $S_i= -2d\Theta_i$, $d\Theta = \Theta-\pi/4$, $c_i = \cos(\theta_i - \varphi_i)$, 
$s_i = \sin(\theta_i - \varphi_i)$.
Eq. (\ref{syseq}) assumes that ground rotational levels of $H^3\Delta_1$ and $C^1\Pi$ states 
can be written according to Eqs. (\ref{Pstate},\ref{Nstate}).
However accounting for interaction with other electronic and rotational states, magnetic field,
modify  Eqs. (\ref{Pstate},\ref{Nstate},\ref{syseq} )
and give rise to systematic errors for other components of $\phi$.
To take into account the perturbation above the numerical calculation was performed.
Following the computational scheme of \cite{Petrov:11, Petrov:14, Petrov:15, Petrov:17c}, wavefunctions of $H$ and $C$ states
in external {\it static} electric and magnetic fields are obtained by numerical diagonalization of the molecular Hamiltonian over the basis set of the electronic-rotational wavefunctions. Detailed features of the Hamiltonian are described in \cite{Petrov:14}. After calculation of wavefunctions the systematic error $\tilde{\phi}$ for precession angle $\phi$ was calculated as 
${\cal A}({\Psi}(\phi=0)) = 2\tilde{P}\tilde{\phi}$. 

\section{Results and discussions}
Table \ref{TResults} gives results of electronic calculations of matrix elements (\ref{dp})-(\ref{gp}).

Comparison of numerical calculations and Eq. (\ref{syseq}) for $\tilde{\phi}^{{\cal {N}}{\cal {E}}}$ is given in Table \ref{res}. 
Typical values of $\Theta_{p,X,Y}$, $\vartheta_{p,X,Y}$, $\theta_{p,X,Y}$ \cite{ACME:17} little deviated from ideal values are used in the Table (\ref{res}).
Calculations show that accounting for perturbations described above does not lead to notable changes in $\tilde{\phi}^{{\cal {N}}{\cal {E}}}$. Numerical calculations are in agreement with
Eq. (\ref{syseq}) within 15\% or less.
Systematic error for the
\eEDM\ due to Stark interference between $E1$ and $M1$ transition amplitudes
is $\tilde{\phi}^{{\cal N}{\cal E}} \sim 10^{-8} rad$. Note that current limits
in terms of $\phi^{{\cal N}{\cal E}}$ are approximately
$\phi^{{\cal N}{\cal E}} < 10^{-5} rad$ for ACME I and 
$\phi^{{\cal N}{\cal E}} < 10^{-6} rad$ for ACME II. 
The systematic error can be further suppressed by about factor of ten
due to the rotation of the readout polarization 
basis by $\theta_{X,Y} \rightarrow \theta_{X,Y} + 90^\circ$ and
a global polarization rotation of both state preparation and readout lasers
$\theta_{p,X,Y} \rightarrow \theta_{p,X,Y} + 90^\circ$ \cite{ACME:17}.
The final systematic error $\tilde{\phi}^{{\cal N}{\cal E}} \sim 10^{-9} rad$ substantially
less than the current limit for $\phi^{{\cal N}{\cal E}}$.
Similar systematic error can be expected for ACME II experiment which
uses $H^3\Delta_1$ to $I^1\Pi$ transition instead of $H^3\Delta_1$ to $C^1\Pi$.

Calculations show that  $\tilde{\phi}^{{\cal N}{\cal B}} \sim 10^{-7} rad$ and $\tilde{\phi}^{{\cal B}} \sim 10^{-8} rad$
which formally are sources of systematic errors in measurement of $g$-factor, $g$-factor difference between the $\Omega$-doublets (see for details Ref. \cite{Petrov:14}), though are several
orders of magnitude less than measured  $\phi^{{\cal N}{\cal B}}$ and $\phi^{{\cal B}}$.

\begin{table*}[p]
\caption{
Calculated electric and magnetic transition dipole moments for the ThO molecule.
}
\label{TResults}
\begin{tabular}{llllll}
        & $ X \rightarrow C $ &        &        &        $ H \rightarrow C $ &             \\
\hline           
\hline
           & $D_+^{XC}$        & $G_{+}^{XC}$    & $J_{+}^{XC}$     & $D_{||}^{HC}$    & $D_{||}^{HC}$      \\
\hline           
20e        & -0.596             & 0.217  & 0.211  & 0.018             & -0.069      \\
OC  & 0.071              & -0.013 & -0.015 & -0.002            & -0.002      \\
\\
Total      & -0.525             & 0.204  & 0.197  & 0.016             & -0.071      \\
\hline
\hline
\end{tabular}
\end{table*}

\begin{table*}[p]
\caption{
Calculated systematic errors $\tilde{\phi}^{{\cal N}{\cal E}}$,  $\tilde{\phi}^{{\cal N}{\cal E}}$, $\tilde{\phi}^{{\cal B}}$ in $10^{-8}rad$ as functions of $\vartheta$, $\varphi$, $\theta$, $\Theta$ (in degrees) for preparation and readout lasers.
Numerical calculations take into account interaction between different rotational
and electronic levels.
}
\label{res}
\begin{tabular}{ l  l  l  l  l  l  l  l l  l  l  l l  l  l  l}
$\vartheta_p$ & $\varphi_p$ & $\theta_p$ & $\Theta_p$ & $\vartheta_X$ & $\varphi_X$ & $\theta_X$ & $\Theta_X$ & 
$\vartheta_Y$ & $\varphi_Y$ & $\theta_Y$ & $\Theta_Y$ & Eq.(\ref{syseq})  &\multicolumn{3}{c}{numerical calculation}\\
            &           &          &          &             &           &          &          &             &           &          &          &  $\tilde{\phi}^{{\cal N}{\cal E}}$    &   $\tilde{\phi}^{{\cal N}{\cal E}}$       &        $\tilde{\phi}^{{\cal N}{\cal B}}$    &   $\tilde{\phi}^{{\cal B}}$    \\
\hline
0.3 &   20. &   1. &    46. &    0.2  &   20. &    46.  &   44.   &     0. &    0. &   136. &   45. &   0.3426 &   0.3423 &  3.621 &  0.4961  \\
0.3 &   20. &   1. &    46. &    0.2  &   20. &    46.  &   44.   &     0. &    0. &   136. &   47. &  -0.1347 &  -0.1353 &  10.0857 & 1.488  \\
0.3 &   20. &   1. &    46. &    0.2  &   20. &    46.  &   44.   &     0. &    0. &   136. &   44. &   0.8199 &   0.8162 & -3.616 & -0.4956  \\
0.3 &   20. &   1. &    46. &    0.2  &   20. &    46.  &   44.   &    0.3 &    0. &   136. &   44. &   0.4596 &   0.4578 & -3.616 & -0.4956  \\
0.3 &   20. &   1. &    46. &    0.2  &   20. &    44.  &   44.   &    0.3 &    0. &   136. &   44. &   0.4765 &   0.4496 & -3.611 & -0.4949  \\
0.3 &   20. &   1. &    46. &    0.2  &   20. &    44.  &   44.   &    0.3 &    0. &   134. &   44. &   0.4224 &   0.4048 & -3.598 & -0.4931  \\
0.3 &   20. &   1. &    46. &    0.2  &   20. &    44.  &   44.   &    0.3 &    0. &   134. &   46. &  -0.2530 &  -0.2978 &  7.207 &  0.9877  \\
0.3 &   20. &   1. &    46. &    0.2  &   20. &    44.  &   44.   &    0.3 &   10. &   134. &   46. &  -0.0933 &  -0.1477 &  7.207 &  0.9877  \\
0.3 &   20. &   1. &    46. &    0.2  &   20. &    44.  &   44.   &    0.3 &  -10. &   134. &   46. &  -0.3675 &  -0.4029 &  7.207 &  0.9877  \\
0.3 &   20. &   1. &    46. &    0.3  &   20. &    44.  &   44.   &    0.3 &  -10. &   134. &   46. &  -0.3835 &  -0.4322 &  7.207 &  0.9877  \\
\hline
\end{tabular}
\end{table*}

\section{Conclusion}
We have calculated the Systematic error for 
the
\eEDM\ search experiment due to Stark interference between $E1$ and $M1$ transition amplitudes. We found that the error is about three
orders of magnitude less than the current limit for 
the
\eEDM\ obtained in ACME II experiment
\cite{ACME:18}


The work is supported by the Russian Science Foundation grant
No. 18-12-00227.


\begin{thebibliography}{26}
\expandafter\ifx\csname natexlab\endcsname\relax\def\natexlab#1{#1}\fi
\expandafter\ifx\csname bibnamefont\endcsname\relax
  \def\bibnamefont#1{#1}\fi
\expandafter\ifx\csname bibfnamefont\endcsname\relax
  \def\bibfnamefont#1{#1}\fi
\expandafter\ifx\csname citenamefont\endcsname\relax
  \def\citenamefont#1{#1}\fi
\expandafter\ifx\csname url\endcsname\relax
  \def\url#1{\texttt{#1}}\fi
\expandafter\ifx\csname urlprefix\endcsname\relax\def\urlprefix{URL }\fi
\providecommand{\bibinfo}[2]{#2}
\providecommand{\eprint}[2][]{\url{#2}}

\bibitem[{\citenamefont{Andreev et~al.}(2018)\citenamefont{Andreev, Ang,
  DeMille, Doyle, Gabrielse, Haefner, Hutzler, Lasner, Meisenhelder, O'Leary
  et~al.}}]{ACME:18}
\bibinfo{author}{\bibfnamefont{V.}~\bibnamefont{Andreev}},
  \bibinfo{author}{\bibfnamefont{D.~G.} \bibnamefont{Ang}},
  \bibinfo{author}{\bibfnamefont{D.}~\bibnamefont{DeMille}},
  \bibinfo{author}{\bibfnamefont{J.~M.} \bibnamefont{Doyle}},
  \bibinfo{author}{\bibfnamefont{G.}~\bibnamefont{Gabrielse}},
  \bibinfo{author}{\bibfnamefont{J.}~\bibnamefont{Haefner}},
  \bibinfo{author}{\bibfnamefont{N.~R.} \bibnamefont{Hutzler}},
  \bibinfo{author}{\bibfnamefont{Z.}~\bibnamefont{Lasner}},
  \bibinfo{author}{\bibfnamefont{C.}~\bibnamefont{Meisenhelder}},
  \bibinfo{author}{\bibfnamefont{B.~R.} \bibnamefont{O'Leary}},
  \bibnamefont{et~al.}, \bibinfo{journal}{Nature}
  \textbf{\bibinfo{volume}{562}}, \bibinfo{pages}{355} (\bibinfo{year}{2018}),
  ISSN \bibinfo{issn}{1476-4687},
  \urlprefix\url{https://doi.org/10.1038/s41586-018-0599-8}.

\bibitem[{\citenamefont{DeMille et~al.}(2001)\citenamefont{DeMille, Bay,
  Bickman, Kawall, Hunter, Krause, Maxwell, and Ulmer}}]{DeMille2001}
\bibinfo{author}{\bibfnamefont{D.}~\bibnamefont{DeMille}},
  \bibinfo{author}{\bibfnamefont{F.}~\bibnamefont{Bay}},
  \bibinfo{author}{\bibfnamefont{S.}~\bibnamefont{Bickman}},
  \bibinfo{author}{\bibfnamefont{D.}~\bibnamefont{Kawall}},
  \bibinfo{author}{\bibfnamefont{L.}~\bibnamefont{Hunter}},
  \bibinfo{author}{\bibfnamefont{D.}~\bibnamefont{Krause}},
  \bibinfo{author}{\bibfnamefont{S.}~\bibnamefont{Maxwell}}, \bibnamefont{and}
  \bibinfo{author}{\bibfnamefont{K.}~\bibnamefont{Ulmer}}, in
  \emph{\bibinfo{booktitle}{AIP Conference Proceedings}}
  (\bibinfo{publisher}{AIP}, \bibinfo{year}{2001}), vol. \bibinfo{volume}{596},
  pp. \bibinfo{pages}{72--83}, ISSN \bibinfo{issn}{0094243X},
  \urlprefix\url{http://link.aip.org/link/?APC/596/72/1&Agg=doi}.

\bibitem[{\citenamefont{Petrov et~al.}(2014)\citenamefont{Petrov, Skripnikov,
  Titov, Hutzler, Hess, O'Leary, Spaun, DeMille, Gabrielse, and
  Doyle}}]{Petrov:14}
\bibinfo{author}{\bibfnamefont{A.~N.} \bibnamefont{Petrov}},
  \bibinfo{author}{\bibfnamefont{L.~V.} \bibnamefont{Skripnikov}},
  \bibinfo{author}{\bibfnamefont{A.~V.} \bibnamefont{Titov}},
  \bibinfo{author}{\bibfnamefont{N.~R.} \bibnamefont{Hutzler}},
  \bibinfo{author}{\bibfnamefont{P.~W.} \bibnamefont{Hess}},
  \bibinfo{author}{\bibfnamefont{B.~R.} \bibnamefont{O'Leary}},
  \bibinfo{author}{\bibfnamefont{B.}~\bibnamefont{Spaun}},
  \bibinfo{author}{\bibfnamefont{D.}~\bibnamefont{DeMille}},
  \bibinfo{author}{\bibfnamefont{G.}~\bibnamefont{Gabrielse}},
  \bibnamefont{and} \bibinfo{author}{\bibfnamefont{J.~M.} \bibnamefont{Doyle}},
  \bibinfo{journal}{Phys. Rev. A} \textbf{\bibinfo{volume}{89}},
  \bibinfo{pages}{062505} (\bibinfo{year}{2014}).

\bibitem[{\citenamefont{Vutha and DeMille}(2009)}]{Vutha2009}
\bibinfo{author}{\bibfnamefont{A.}~\bibnamefont{Vutha}} \bibnamefont{and}
  \bibinfo{author}{\bibfnamefont{D.}~\bibnamefont{DeMille}},
  \bibinfo{journal}{arXiv}  (\bibinfo{year}{2009}), \eprint{0907.5116},
  \urlprefix\url{http://arxiv.org/abs/0907.5116}.

\bibitem[{\citenamefont{Petrov}(2015)}]{Petrov:15}
\bibinfo{author}{\bibfnamefont{A.~N.} \bibnamefont{Petrov}},
  \bibinfo{journal}{Phys.\ Rev.\ A} \textbf{\bibinfo{volume}{91}},
  \bibinfo{pages}{062509} (\bibinfo{year}{2015}).

\bibitem[{\citenamefont{Petrov}(2018)}]{Petrov:18}
\bibinfo{author}{\bibfnamefont{A.~N.} \bibnamefont{Petrov}},
  \bibinfo{journal}{Phys. Rev. A} \textbf{\bibinfo{volume}{97}},
  \bibinfo{pages}{052504} (\bibinfo{year}{2018}).

\bibitem[{\citenamefont{Baron et~al.}(2014)\citenamefont{Baron, Campbell,
  Demille, Doyle, Gabrielse, Gurevich, Hess, Hutzler, Kirilov, Kozyryev
  et~al.}}]{Baron2013}
\bibinfo{author}{\bibfnamefont{J.}~\bibnamefont{Baron}},
  \bibinfo{author}{\bibfnamefont{W.~C.} \bibnamefont{Campbell}},
  \bibinfo{author}{\bibfnamefont{D.}~\bibnamefont{Demille}},
  \bibinfo{author}{\bibfnamefont{J.~M.} \bibnamefont{Doyle}},
  \bibinfo{author}{\bibfnamefont{G.}~\bibnamefont{Gabrielse}},
  \bibinfo{author}{\bibfnamefont{Y.~V.} \bibnamefont{Gurevich}},
  \bibinfo{author}{\bibfnamefont{P.~W.} \bibnamefont{Hess}},
  \bibinfo{author}{\bibfnamefont{N.~R.} \bibnamefont{Hutzler}},
  \bibinfo{author}{\bibfnamefont{E.}~\bibnamefont{Kirilov}},
  \bibinfo{author}{\bibfnamefont{I.}~\bibnamefont{Kozyryev}},
  \bibnamefont{et~al.}, \bibinfo{journal}{Science}
  \textbf{\bibinfo{volume}{343}}, \bibinfo{pages}{269} (\bibinfo{year}{2014}),
  ISSN \bibinfo{issn}{1095-9203}

\bibitem[{\citenamefont{Baron et~al.}(2017)\citenamefont{Baron, Campbell,
  DeMille, Doyle, Gabrielse, Gurevich, Hess, Hutzler, Kirilov, Kozyryev
  et~al.}}]{ACME:17}
\bibinfo{author}{\bibfnamefont{J.}~\bibnamefont{Baron}},
  \bibinfo{author}{\bibfnamefont{W.~C.} \bibnamefont{Campbell}},
  \bibinfo{author}{\bibfnamefont{D.}~\bibnamefont{DeMille}},
  \bibinfo{author}{\bibfnamefont{J.~M.} \bibnamefont{Doyle}},
  \bibinfo{author}{\bibfnamefont{G.}~\bibnamefont{Gabrielse}},
  \bibinfo{author}{\bibfnamefont{Y.~V.} \bibnamefont{Gurevich}},
  \bibinfo{author}{\bibfnamefont{P.~W.} \bibnamefont{Hess}},
  \bibinfo{author}{\bibfnamefont{N.~R.} \bibnamefont{Hutzler}},
  \bibinfo{author}{\bibfnamefont{E.}~\bibnamefont{Kirilov}},
  \bibinfo{author}{\bibfnamefont{I.}~\bibnamefont{Kozyryev}},
  \bibnamefont{et~al.}, \bibinfo{journal}{New Journal of Physics}
  \textbf{\bibinfo{volume}{19}}, \bibinfo{pages}{073029}
  (\bibinfo{year}{2017}),
  \urlprefix\url{http://stacks.iop.org/1367-2630/19/i=7/a=073029}.

\bibitem[{\citenamefont{Skripnikov
  et~al.}(2013{\natexlab{a}})\citenamefont{Skripnikov, Petrov, and
  Titov}}]{Skripnikov:13c}
\bibinfo{author}{\bibfnamefont{L.~V.} \bibnamefont{Skripnikov}},
  \bibinfo{author}{\bibfnamefont{A.~N.} \bibnamefont{Petrov}},
  \bibnamefont{and} \bibinfo{author}{\bibfnamefont{A.~V.} \bibnamefont{Titov}},
  \bibinfo{journal}{J.\ Chem.\ Phys.} \textbf{\bibinfo{volume}{139}},
  \bibinfo{eid}{221103} (\bibinfo{year}{2013}{\natexlab{a}}).

\bibitem[{\citenamefont{Skripnikov and
  Titov}(2015{\natexlab{a}})}]{Skripnikov:15a}
\bibinfo{author}{\bibfnamefont{L.~V.} \bibnamefont{Skripnikov}}
  \bibnamefont{and} \bibinfo{author}{\bibfnamefont{A.~V.} \bibnamefont{Titov}},
  \bibinfo{journal}{J.\ Chem.\ Phys.} \textbf{\bibinfo{volume}{142}},
  \bibinfo{eid}{024301} (\bibinfo{year}{2015}{\natexlab{a}}).

\bibitem[{\citenamefont{Skripnikov}(2016)}]{Skripnikov:16b}
\bibinfo{author}{\bibfnamefont{L.~V.} \bibnamefont{Skripnikov}},
  \bibinfo{journal}{J.\ Chem.\ Phys.} \textbf{\bibinfo{volume}{145}},
  \bibinfo{pages}{214301} (\bibinfo{year}{2016}).

\bibitem[{\citenamefont{Denis and Fleig}(2016)}]{Fleig:16}
\bibinfo{author}{\bibfnamefont{M.}~\bibnamefont{Denis}} \bibnamefont{and}
  \bibinfo{author}{\bibfnamefont{T.}~\bibnamefont{Fleig}},
  \bibinfo{journal}{The Journal of Chemical Physics}
  \textbf{\bibinfo{volume}{145}}, \bibinfo{pages}{214307}
  (\bibinfo{year}{2016}).

\bibitem[{\citenamefont{{Dunning, Jr}}(1989)}]{Dunning:89}
\bibinfo{author}{\bibfnamefont{T.~H.} \bibnamefont{{Dunning, Jr}}},
  \bibinfo{journal}{J.\ Chem.\ Phys.} \textbf{\bibinfo{volume}{90}},
  \bibinfo{pages}{1007} (\bibinfo{year}{1989}).

\bibitem[{\citenamefont{Kendall et~al.}(1992)\citenamefont{Kendall, {Dunning,
  Jr}, and Harrison}}]{Kendall:92}
\bibinfo{author}{\bibfnamefont{R.~A.} \bibnamefont{Kendall}},
  \bibinfo{author}{\bibfnamefont{T.~H.} \bibnamefont{{Dunning, Jr}}},
  \bibnamefont{and} \bibinfo{author}{\bibfnamefont{R.~J.}
  \bibnamefont{Harrison}}, \bibinfo{journal}{J.\ Chem.\ Phys.}
  \textbf{\bibinfo{volume}{96}}, \bibinfo{pages}{6796} (\bibinfo{year}{1992}).

\bibitem[{\citenamefont{Skripnikov
  et~al.}(2013{\natexlab{b}})\citenamefont{Skripnikov, Mosyagin, and
  Titov}}]{Skripnikov:13a}
\bibinfo{author}{\bibfnamefont{L.~V.} \bibnamefont{Skripnikov}},
  \bibinfo{author}{\bibfnamefont{N.~S.} \bibnamefont{Mosyagin}},
  \bibnamefont{and} \bibinfo{author}{\bibfnamefont{A.~V.} \bibnamefont{Titov}},
  \bibinfo{journal}{Chem.\ Phys.\ Lett.} \textbf{\bibinfo{volume}{555}},
  \bibinfo{pages}{79} (\bibinfo{year}{2013}{\natexlab{b}}).

\bibitem[{\citenamefont{Mosyagin et~al.}(2010)\citenamefont{Mosyagin,
  Zaitsevskii, and Titov}}]{Mosyagin:10a}
\bibinfo{author}{\bibfnamefont{N.~S.} \bibnamefont{Mosyagin}},
  \bibinfo{author}{\bibfnamefont{A.~V.} \bibnamefont{Zaitsevskii}},
  \bibnamefont{and} \bibinfo{author}{\bibfnamefont{A.~V.} \bibnamefont{Titov}},
  \bibinfo{journal}{Review of Atomic and Molecular Physics}
  \textbf{\bibinfo{volume}{1}}, \bibinfo{pages}{63} (\bibinfo{year}{2010}).

\bibitem[{\citenamefont{Mosyagin et~al.}(2016)\citenamefont{Mosyagin,
  Zaitsevskii, Skripnikov, and Titov}}]{Mosyagin:16}
\bibinfo{author}{\bibfnamefont{N.~S.} \bibnamefont{Mosyagin}},
  \bibinfo{author}{\bibfnamefont{A.~V.} \bibnamefont{Zaitsevskii}},
  \bibinfo{author}{\bibfnamefont{L.~V.} \bibnamefont{Skripnikov}},
  \bibnamefont{and} \bibinfo{author}{\bibfnamefont{A.~V.} \bibnamefont{Titov}},
  \bibinfo{journal}{Int.\ J.\ Quantum Chem.} \textbf{\bibinfo{volume}{116}},
  \bibinfo{pages}{301} (\bibinfo{year}{2016}), ISSN \bibinfo{issn}{1097-461X}.

\bibitem[{\citenamefont{K\'{a}llay and Gauss}(2004)}]{Kallay:5}
\bibinfo{author}{\bibfnamefont{M.}~\bibnamefont{K\'{a}llay}} \bibnamefont{and}
  \bibinfo{author}{\bibfnamefont{J.}~\bibnamefont{Gauss}},
  \bibinfo{journal}{J.\ Chem.\ Phys.} \textbf{\bibinfo{volume}{121}},
  \bibinfo{pages}{9257} (\bibinfo{year}{2004}).

\bibitem[{DIR()}]{DIRAC15}
\bibinfo{note}{DIRAC, a relativistic ab initio electronic structure program,
  Release DIRAC15 (2015), written by R. Bast, T. Saue, L. Visscher, and H. J.
  Aa. Jensen, with contributions from V. Bakken, K. G. Dyall, S. Dubillard, U.
  Ekstroem, E. Eliav, T. Enevoldsen, E. Fasshauer, T. Fleig, O. Fossgaard, A.
  S. P. Gomes, T. Helgaker, J. Henriksson, M. Ilias, Ch. R. Jacob, S. Knecht,
  S. Komorovsky, O. Kullie, J. K. Laerdahl, C. V. Larsen, Y. S. Lee, H. S.
  Nataraj, M. K. Nayak, P. Norman, G. Olejniczak, J. Olsen, Y. C. Park, J. K.
  Pedersen, M. Pernpointner, R. Di Remigio, K. Ruud, P. Salek, B.
  Schimmelpfennig, J. Sikkema, A. J. Thorvaldsen, J. Thyssen, J. van Stralen,
  S. Villaume, O. Visser, T. Winther, and S. Yamamoto (see
  http://www.diracprogram.org).}

\bibitem[{MRC()}]{MRCC2013}
\bibinfo{note}{{\sc mrcc}, a quantum chemical program suite written by M.
  K\'{a}llay, Z. Rolik, I. Ladj\'{a}nszki, L. Szegedy, B. Lad\'{o}czki, J.
  Csontos, and B. Kornis. See also Z. Rolik and M. K\'{a}llay, J. Chem. Phys.
  135, 104111 (2011), as well as: www.mrcc.hu}.

\bibitem[{\citenamefont{Skripnikov and
  Titov}(2015{\natexlab{b}})}]{Skripnikov:15b}
\bibinfo{author}{\bibfnamefont{L.~V.} \bibnamefont{Skripnikov}}
  \bibnamefont{and} \bibinfo{author}{\bibfnamefont{A.~V.} \bibnamefont{Titov}},
  \bibinfo{journal}{Phys. Rev. A} \textbf{\bibinfo{volume}{91}},
  \bibinfo{pages}{042504} (\bibinfo{year}{2015}{\natexlab{b}}).

\bibitem[{\citenamefont{Skripnikov et~al.}(2015)\citenamefont{Skripnikov,
  Petrov, Titov, Mawhorter, Baum, Sears, and Grabow}}]{Skripnikov:15d}
\bibinfo{author}{\bibfnamefont{L.~V.} \bibnamefont{Skripnikov}},
  \bibinfo{author}{\bibfnamefont{A.~N.} \bibnamefont{Petrov}},
  \bibinfo{author}{\bibfnamefont{A.~V.} \bibnamefont{Titov}},
  \bibinfo{author}{\bibfnamefont{R.~J.} \bibnamefont{Mawhorter}},
  \bibinfo{author}{\bibfnamefont{A.~L.} \bibnamefont{Baum}},
  \bibinfo{author}{\bibfnamefont{T.~J.} \bibnamefont{Sears}}, \bibnamefont{and}
  \bibinfo{author}{\bibfnamefont{J.-U.} \bibnamefont{Grabow}},
  \bibinfo{journal}{Phys. Rev. A} \textbf{\bibinfo{volume}{92}},
  \bibinfo{pages}{032508} (\bibinfo{year}{2015}).

\bibitem[{\citenamefont{Vutha et~al.}(2011)\citenamefont{Vutha, Spaun,
  Gurevich, Hutzler, Kirilov, Doyle, Gabrielse, and DeMille}}]{Vutha2011}
\bibinfo{author}{\bibfnamefont{A.~C.} \bibnamefont{Vutha}},
  \bibinfo{author}{\bibfnamefont{B.}~\bibnamefont{Spaun}},
  \bibinfo{author}{\bibfnamefont{Y.~V.} \bibnamefont{Gurevich}},
  \bibinfo{author}{\bibfnamefont{N.~R.} \bibnamefont{Hutzler}},
  \bibinfo{author}{\bibfnamefont{E.}~\bibnamefont{Kirilov}},
  \bibinfo{author}{\bibfnamefont{J.~M.} \bibnamefont{Doyle}},
  \bibinfo{author}{\bibfnamefont{G.}~\bibnamefont{Gabrielse}},
  \bibnamefont{and} \bibinfo{author}{\bibfnamefont{D.}~\bibnamefont{DeMille}},
  \bibinfo{journal}{Physical Review A} \textbf{\bibinfo{volume}{84}},
  \bibinfo{pages}{034502} (\bibinfo{year}{2011}), ISSN
  \bibinfo{issn}{1050-2947},
  \urlprefix\url{http://link.aps.org/doi/10.1103/PhysRevA.84.034502}.

\bibitem[{\citenamefont{Hess}(2014)}]{Hess2014Thesis}
\bibinfo{author}{\bibfnamefont{P.~W.} \bibnamefont{Hess}}, Ph.D. thesis,
  \bibinfo{school}{Harvard University} (\bibinfo{year}{2014}).

\bibitem[{\citenamefont{Petrov}(2011)}]{Petrov:11}
\bibinfo{author}{\bibfnamefont{A.~N.} \bibnamefont{Petrov}},
  \bibinfo{journal}{Phys.\ Rev.\ A} \textbf{\bibinfo{volume}{83}},
  \bibinfo{pages}{024502} (\bibinfo{year}{2011}).

\bibitem[{\citenamefont{Petrov}(2017)}]{Petrov:17c}
\bibinfo{author}{\bibfnamefont{A.~N.} \bibnamefont{Petrov}},
  \bibinfo{journal}{Phys. Rev. A} \textbf{\bibinfo{volume}{95}},
  \bibinfo{pages}{062501} (\bibinfo{year}{2017}),
  \urlprefix\url{https://link.aps.org/doi/10.1103/PhysRevA.95.062501}.

\end{thebibliography}

\end{document}